\documentclass[aps,a4paper,11pt,reprint]{revtex4-2}
\usepackage{xr}
\usepackage{times}
\usepackage{graphicx}
\usepackage{amsmath}
\usepackage{amssymb}
\usepackage{booktabs}
\usepackage{multirow}
\usepackage{bm}
\usepackage{siunitx}

\usepackage{hyperref}
\usepackage{cleveref}

\begin{document}
	
\title{Competition between vacancy creation and filling \\ in defect-engineering of hBN}
	
\author
{Shrirang Chokappa$^{1,2,\dag,*}$, Manuel Längle$^{1,3,\dag,*,}$, Barbara Maria Mayer$^{1,\ddagger}$, Vladimir Zoba\v{c}$^{1}$, Jacob Madsen$^1$, Diana Propst$^{1,2}$, David Lamprecht$^{1,4}$, Philipp Irschik$^{1,2}$, Arixin Bo$^1$, Clara Kofler$^{1,2}$, Vinzent Hana$^{1,\ddagger}$, Fabian Kraft$^1$, Clemens Mangler$^1$, Lado Filipovic$^{4}$, Toma Susi$^1$, Jani Kotakoski$^{1,*}$\\
	\footnotesize{$^1$University of Vienna, Faculty of Physics,\\ Boltzmanngasse 5, 1090 Vienna, Austria}\\
	\footnotesize{$^2$University of Vienna, Vienna Doctoral School in Physics,\\ Boltzmanngasse 5, 1090 Vienna, Austria}\\
	\footnotesize{$^3$Univ. Paris-Saclay, CNRS, Laboratoire de Physique des Solides,\\ 91405, Orsay, France}\\
	\footnotesize{$^4$Institute for Microelectronics, TU Wien,\\Gußhausstraße 27-29 / E360, 1040 Vienna, Austria}\\
	\footnotesize{$^\dag$ These authors contributed equally to this work.}\\
	\footnotesize{$^*$ Email: shrirang.chokappa@univie.ac.at, manuel.langle@universite-paris-saclay.fr, jani.kotakoski@univie.ac.at\\}
	\footnotesize{$\ddagger$ Current affiliation: (BMM) Materials Physics, Department of Physics and Astronomy, Uppsala University, Uppsala, Sweden; (VH) University of Glasgow, School of Physics and Astronomy, Glasgow G12 8QQ}
}
	
\date{September 10, 2026}
	
	\begin{abstract}
		\begin{center}
			\textbf{ABSTRACT}
		\end{center}
		Hexagonal boron nitride (hBN) has recently become the focus of intense research as a material that can host quantum emitters.
		It is known that such emission is related to point defects, but in order to conclusively correlate specific defects to their spectra, having control over the defect creation mechanism is required.
		Here, we prepare freestanding, monolayer hBN samples and irradiate them with ultra-low-energy (150~eV) Ar$^+$ ions.
		The samples are characterized before and after irradiation via scanning transmission electron microscopy to assess the defect density and distribution.
		Contrary to what analytical potential molecular dynamics simulations have predicted, we predominantly observe boron single vacancies after ion irradiation, followed by double vacancies at half the count.
		Moreover, we also observe that vacancy filling with Si and C impurity atoms plays a more significant role in the created defects than previously assumed, potentially posing a problem for selective creation of quantum emitters in hBN.
	\end{abstract}
	
	\maketitle	
	\section*{Introduction}
	
	Although hexagonal boron nitride (hBN) is mostly known as the electrically insulating counterpart to graphene, it has recently been receiving increasing attention as a solid state host for quantum emitters~\cite{sajid_Singlephoton_2020}.
	Quantum emitters in hBN exhibit good stability over a wide temperature range~\cite{kianinia_Robust_2017}, as well as bright emission in the zero-phonon line~\cite{tran_Quantum_2016}.
	Combined with their wide spectral range~\cite{tran_Robust_2016, bourrellier_Bright_2016, gottscholl_Initialization_2020}, due to there being multiple different emitter types, and the possibility to get lifetime-limited emission at room temperature~\cite{hoese_Mechanical_2020,dietrich_Solidstate_2020}, hBN has proven to be an attractive material for several advanced applications~\cite{deleon_Materials_2021}. 
	Although it is well known that quantum emitters are associated with point defects in the material~\cite{tran_Robust_2016,bourrellier_Bright_2016,tawfik_Firstprinciples_2017,su_Tuning_2022,wong_Characterization_2015,abdi_Color_2018}, so far no satisfactory direct correlation between the defect structures and quantum emission properties has been established \cite{baber_Excited_2022,fischer_Combining_2023,fournier_Positioncontrolled_2021,gale_SiteSpecific_2022,mackoit-sinkeviciene_Carbon_2019}.
	Therefore, being able to selectively create specific types of point defects in hBN and to characterize their atomic structure would be desirable.
	It has already been shown that different kinds of irradiation can be used to create quantum emitters in hBN, including lasers~\cite{gan_LargeScale_2022}, ions~\cite{choi_Engineering_2016}, neutrons~\cite{zhang_Discrete_2019} and electrons~\cite{exarhos_Optical_2017,su_Tuning_2022}.
	So far, only in the case of electrons in monolayer hBN~\cite{bui_Creation_2023, javed_Origin_2026a}, and high energy ions (500~kV He$^{\mathrm{+}}$) for multilayer hBN~\cite{liang_SiteSelective_2025} has a direct correlation between irradiation and the exact atomic structure been established, partially due to the challenges related to defect classification for non-monolayer structures~\cite{lamprecht_Single_2026}.
	For large-scale applications, a method to generate defects more effectively would be necessary.
	Electron irradiation, particularly in scanning transmission electron microscopy (STEM), has the disadvantage of being extremely localized, while high-energy ion irradiation efficiency for defect creation is relatively low, especially in thin specimens.
	
	Analytical potential molecular dynamics simulations have shown that low-energy noble-gas ions should have high defect creation efficiencies.
	According to simulations with a potential benchmarked against density functional theory (DFT)~\cite{ghaderzadeh_Atomistic_2021}, the best selectivity should be achievable at ion energies lower than 200~eV, where nitrogen single vacancies were reported to be the most prominent defect with noble gas projectiles heavier than Ne.
	In an earlier work, Lehtinen et al.~\cite{lehtinen_Production_2011} predicted that about 0.8 defects are created per 100~eV Ar ion with roughly equal probabilities ($\sim$0.2) for creating boron ($\mathrm{V_B}$) and nitrogen single vacancies ($\mathrm{V_N}$) and roughly twice that for double vacancies ($\mathrm{V_{BN}}$) in monolayer hBN.
	Ghaderzadeh et al.~\cite{ghaderzadeh_Atomistic_2021}, on the other hand, reported about 0.2 defects being created per 150~eV Ar ion at low fluences.
	They predict a higher density of $\mathrm{V_N}$ ($\sim$0.4) than for $\mathrm{V_{BN}}$  ($\sim$0.3) and $\mathrm{V_B}$ ($\sim$0.2).
	However, these predictions have not been confirmed experimentally.
	
	For electron irradiation, it was experimentally shown that intermediate electron energies (60--100~keV) are more likely to displace boron than nitrogen~\cite{bui_Creation_2023}.
	Electron irradiation in a low-pressure oxygen atmosphere has also been shown to result in preferential removal of boron, at least from pore edges~\cite{javed_Origin_2026a}, often leading to nitrogen-terminated pores.
	Whether these effects play a role also for ultra-low-energy singly-charged ion irradiation of hBN remains unclear.
	However, they certainly do play a role when quantifying ion irradiation effects using electron microscopy in non-ultra-high vacuum conditions.
	Recent reports of atomic-level (S)TEM characterisation of ion-induced defects in hBN for mono-~\cite{byrne_Atomic_2025} and multilayers~\cite{liang_SiteSelective_2025}, find predominantly point defects being created.
	Byrne et al.~\cite{byrne_Atomic_2025} used 25~keV He${^+}$ ions with a fluence of 75 ions~nm$^{-2}$, and found a defect density of 0.16~nm$^{-2}$, with 80~kV TEM imaging at a dose rate ranging from 500 to 700~e$^{-}$/\AA$^{2}$/s. 
	Without irradiation, a baseline defect density of 0.03~nm$^{-2}$ was measured.
	Liang et al.~\cite{liang_SiteSelective_2025} used 500~keV He${^+}$ with a fluence of 30 ions~nm$^{-2}$ on multilayer hBN and measured a total of 0.08 defects~nm$^{-2}$ with high angle annular dark field (HAADF)-STEM imaging at 80~kV, which were predominately boron single vacancies and its substitutions by impurity atoms.
	A study conducted by Wang et al.~\cite{wang_Photoluminescence_2018} characterized defects in exfoliated monolayer hBN at the atomic scale; however, no ion irradiation was conducted here.
	An unexpectedly high intrinsic defect density of 0.14~nm$^{-2}$ was reported, using ADF-STEM imaging at 60~kV.
	
	These results show that large-scale atomic-resolution analysis of ion-induced defects in hBN using electron microscopy is necessary for their characterisation, taking into account the effect of electron-beam damage~\cite{bui_Creation_2023} and beam-assisted chemical etching~\cite{javed_Origin_2026a}.
	However, it should be noted that exact defect identification using ADF imaging for few- and multilayer hBN is beyond practical feasibility as intensity differences between boron or nitrogen dominated columns (AA$'$ stacking) and defects become increasingly indistinguishable with increasing thickness of hBN, which is further complicated by common residual aberrations~\cite{lamprecht_Single_2026}.
	So far the only direct experimental comparison to the simulation studies~\cite{ghaderzadeh_Atomistic_2021,lehtinen_Production_2011} for low-energy ion irradiation of monolayer hBN was carried out in Refs.~\cite{propst_Automated_2024,langle_Defectengineering_2024}.
	Although ion irradiation did yield some selectivity in the defects created, the results were unclear because of a bimodal ion-beam energy profile and, importantly, the imaging conditions failed to conclusively resolve impurity atoms in the lattice.
	
	Here, we improve on this work to make the comparison between experiments and simulations more straightforward.
	Specifically, we conduct singly charged argon irradiation experiments at 150~eV with a better defined ion energy and improved STEM image acquisition parameters.
	As energies even up to 40~keV are considered low in the ion beam community~\cite{wilhelm_Unraveling_2019}, we term the irradiation used here ultra-low-energy to avoid confusion.
	We prepare freestanding hBN samples from commercial monolayer hBN grown via chemical vapor deposition (CVD) through electrochemical delamination and heat the samples in ultra-high-vacuum (UHV) as in Ref.~\cite{irschik_Atomically_2026} to reveal large areas of clean, monolayer hBN.
	The samples are irradiated using ca. 150~eV Ar${^+}$ ions generated from a microwave plasma source.
	The defect densities and distributions were measured using semi-automated ADF-STEM imaging at 60~kV in ultra-high vacuum, and are analysed using a convolutional neural network (CNN)-assisted defect identification workflow~\cite{trentino_AtomicLevel_2021,propst_Automated_2024}.
	We find an intrinsic, pre-irradiation defect density of 0.021~nm$^{-2}$, similar to what has been reported in literature for CVD-grown monolayer hBN~\cite{byrne_Atomic_2025}.
	The total defect density after irradiation was measured to be 0.228~nm$^{-2}$, for an average ion fluence of 0.8$\pm$0.2~ ions~nm$^{-2}$.
	This defect density is sufficiently high to confirm irradiation-related damage yet low enough that most ions encounter pristine hBN, ensuring that ions rarely impact the exact same location more than once.
	The most prevalent defect is the boron single vacancy ($\mathrm{V_B}$) with a defect density of 0.089~nm$^{-2}$, followed by double vacancies at 0.042~nm$^{-2}$. Interestingly, these results do not align with the predicted defect distributions reported by the simulation studies~\cite{ghaderzadeh_Atomistic_2021,lehtinen_Production_2011}.
	
	We also, unexpectedly, observe a large number of silicon atom substitutions in boron sites ($\mathrm{Si_B}$) at a density of 0.035~nm$^{-2}$.
	This vacancy filling appears to be an uncontrollable by-product of ion irradiation.
	Contrary to the limited number of impurity atom substitutions reported in Refs.~\cite{propst_Automated_2024,langle_Defectengineering_2024}, these post-irradiation impurity atoms predominantly occupy boron sites, hence ruling out the possibility that the discrepancy between simulations and experiments could be explained by vacancy filling that obscures the number of nitrogen single vacancies created by the irradiation.
	Similar ion irradiation experiments conducted on graphene~\cite{trentino_AtomicLevel_2021,joudi_CorrugationDominated_2025} yielded primarily double vacancies and much lower vacancy filling by impurity atoms, indicating that a different mechanism could be taking place in hBN.
	Overall, the results presented here show that a degree of defect selectivity can indeed be achieved in hBN with ultra-low-energy noble gas irradiation, despite a competing mechanism vacancy filling by impurity atoms.
	This also confirms that defect production is much more efficient with $<$1~keV ions compared to $>$20~keV ions, as would be expected from the larger displacement cross section at lower energies.
	
	\section*{Results}
	
	\subsection*{Sample preparation and characterisation}
	The samples were prepared using the electrochemical delamination method~\cite{wang_Electrochemical_2011, irschik_Atomically_2026} with polymethyl methacrylate (PMMA) assisted transfer of commercially available CVD-grown monolayer hBN on copper foil (Sigma-Aldrich) onto Quantifoil (QF) gold (Au) TEM grids with an amorphous carbon support membrane.
	The samples were placed in a warm acetone bath followed by room temperature isopropyl alcohol (IPA), for 1 h each, to dissolve the PMMA and remove solvent residue.
	After preparation, the samples were inserted into the vacuum system~\cite{mangler_Materials_2022} that connects all instruments used in this study through ultra-high-vacuum transfer lines (base pressure typically in the 10$^{-9}$~mbar regime).
	Upon insertion, the samples undergo a routine bake at 150$^{\circ}$C for ca. 10 h in vacuum to remove water and weakly adsorbed contamination.
	To further remove more strongly bound surface contamination, the samples were heated in UHV at 250$^{\circ}$C for 180~min~\cite{irschik_Atomically_2026}.
	This process is able to completely remove contamination from pristine monolayer regions but not from defects, metal nanoclusters, grain boundaries, and multilayer regions.
	The samples are then surveyed via STEM using medium-angle annular dark field (MAADF) imaging with the Nion UltraSTEM 100 at 60~kV in UHV.
	
	\begin{figure}[t]
		\centering
		\includegraphics[width=1\linewidth]{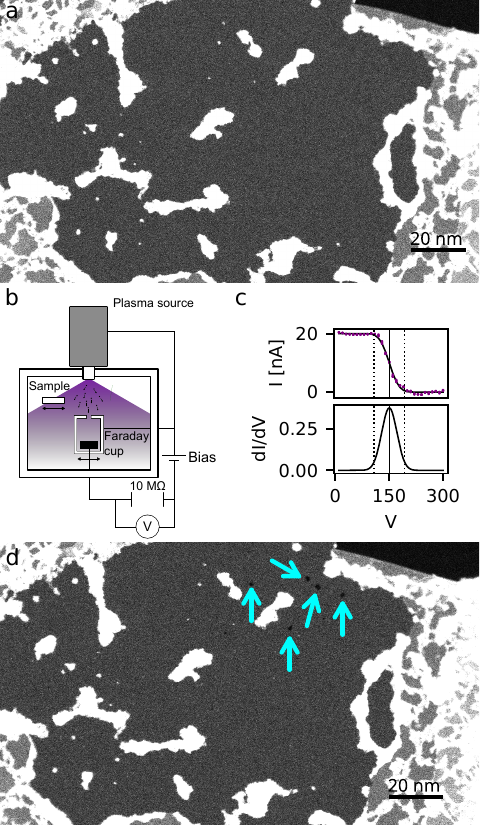}
		\caption{{\bf Sample cleaning, plasma set-up and beam energy profile.}
			(a) MAADF-STEM image of heat treated (250$^{\circ}$C/180~min) freestanding, monolayer hBN before irradiation.
			The darkest contrast corresponds to vacuum, the second-darkest to monolayer hBN, and all brighter features are hydrocarbon or metal contamination, or multilayers.
			(b) Schematic representation of the plasma irradiation setup and Faraday cup used to measure the energy profile.
			(c) Top: Background-subtracted ion current $I$ (purple circles) and fitted beam energy profile ($\mathrm{d}I/\mathrm{d}V$) (black line) as a function of the bias voltage ($V$).
			Bottom: Absolute value of the derivative of the fit representing the energy spread.
			(d) MAADF-STEM image of the same freestanding, monolayer hBN region after irradiation. The cyan arrows indicate larger pores formed by the electron beam during imaging.
		}
		\label{fig:methods_plasma}
	\end{figure}
	
	An example overview image acquired prior to irradiation is shown in Fig.~\ref{fig:methods_plasma}a with ca. 100$\times$100 nm$^2$ monolayer hBN containing a few contamination patches on the surface.
	Supplementary Fig.~S1 shows all the overview images of the regions where data was acquired (pre- and post-irradiation).
	A total area of ca. 15,200~nm$^2$ was imaged for the as-prepared hBN samples using a semi-automatic imaging tool as described later.
	After masking out regions containing contamination or scan distortions, the total valid monolayer area that was identified by the CNN was ca. 11,500~nm$^2$.
	After pre-irradiation characterisation, the samples were transferred in vacuum to a chamber containing a plasma generator, illustrated in Fig.~\ref{fig:methods_plasma}b, which we use as a source for the ultra-low-energy Ar$^+$ ions to create defects, following Ref.~\cite{trentino_AtomicLevel_2021}.

	\subsection*{Ion-irradiation}
	Before and after each irradiation, the profile was measured by biasing the Faraday cup and measuring the current as a function of the bias voltage (see Methods for more details).
	A typical beam current (top) and energy profile (bottom) of the ions is shown in Fig.~\ref{fig:methods_plasma}c.
	The ions had a mean energy between 149 and 155~eV with a standard deviation between 21 and 28~eV over all experiments.
	All energy profiles are shown in Supplementary Fig.~S2, and the corresponding parameters are given in Supplementary Tab.~I.
	Each sample was placed in the ion beam path for 150~s resulting in an average fluence of 0.8$\pm$0.2 ions~nm$^{-2}$ (see Methods for more details).
	The pressure in the chamber varied between 3.6 and 4.2 $\times$10$^{-6}$~mbar depending on the experiment, but remained stable during each irradiation.
	Once irradiated, the samples were imaged once again via STEM, to estimate the post-irradiation defect density and distribution.
	No additional cleaning was necessary after the irradiation, as we saw no significant change in the surface contamination.
	Fig.~\ref{fig:methods_plasma}d is a post-irradiation overview image of the same region as Fig.~\ref{fig:methods_plasma}a.
	The cyan arrows indicate a few larger pores that were formed due to electron beam damage while tuning and thus exposing the sample locally to a high electron dose.

	\begin{figure}[t]
		\centering
		\includegraphics[width=1\linewidth]{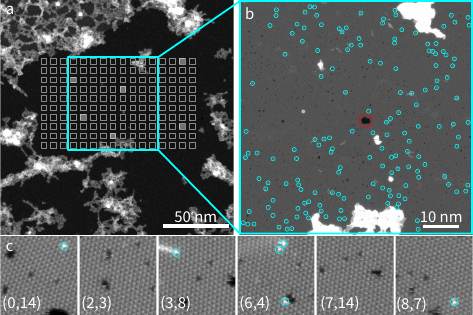}
		\caption{{\bf Semi-automated STEM imaging.}
			(a) STEM-MAADF overview image of a freestanding, thermally cleaned, irradiated sample region.
			The darkest contrast is clean, freestanding hBN and the brighter contrast corresponds to contamination or multilayer regions.
			The gray squares mark planned imaging positions within the user-defined region.
			(b) 64$\times$64~nm$^2$ image of a part of the imaged region where vacancy filling by silicon atoms can be seen, indicated with cyan circles. The red circle indicates a pore created where the electron beam was parked after the semi-automatic imaging.
			(c) Example individual frames from the semi-automated image acquisition, with their corresponding grid coordinates (horizontal, vertical) starting from the top left at 0 and going to the bottom right. Silicon atoms are also indicated with cyan circles here.
			The chosen frames are marked with gray-filled squares in (a) and have a size of 4$\times$4~nm$^2$.
		}
		\label{fig:scanmap}
	\end{figure}
	
	\subsection*{Atomic scale imaging and analysis}
	Clean monolayer regions of ca. 200$\times$200~nm$^2$ were chosen and a semi-automated atomic-resolution imaging tool~\cite{mittelberger_Automated_2017} was used to acquire data as in Ref.~\cite{trentino_AtomicLevel_2021}, in order to estimate the defect density and its distribution.
	Fig.~\ref{fig:scanmap}a shows an example post-irradiation area that was used for semi-automated imaging.
	With a pre-tuned electron beam, the focus was set at four corners of a large rectangular area, with a typical size of 150$\times$150~nm$^2$.
	The stage moves through the defined area collecting hundreds of smaller (4$\times$4~nm$^2$) images at atomic resolution with a dwell time of 16~$\mathrm{\mu}$s, 512$\times$512 px, an offset of 6~nm between images (to avoid overlapping frames), and a sleep time of 3~s (to avoid stage-drift artifacts in the acquired images).
	The small gray squares in Fig.~\ref{fig:scanmap}a indicate planned positions where the smaller FOV images were taken from.
	Due to hysteresis of the stage, in reality the acquisition positions are slightly displaced from these planned positions~\cite{mittelberger_Automated_2017}.
	This tool allows the collection of a large number of images while minimizing the electron dose that the sample is exposed to during acquisition.
	Fig.~\ref{fig:scanmap}b is a magnified image at lattice resolution showing monolayer hBN containing a high number of defects, with silicon atom substitutions circled in cyan.
	The raw images of a few highlighted squares in Fig.~\ref{fig:scanmap}a are shown in Fig.~\ref{fig:scanmap}c with their respective grid coordinates.
	Multiple datasets were collected on the samples both before and after ultra-low-energy Ar$^{+}$ irradiation using this method, with their respective overview images shown in Supplementary Fig.~S1.
	
	\begin{figure}[t]
		\centering
		\includegraphics[width=1\linewidth]{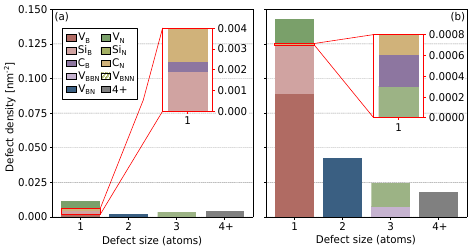}
		\caption{{\bf Defect density and distribution.}
			(a) Intrinsic defect density as a function of the defect size (number of missing atoms).
			(b) Post-irradiation defect density as a function of the defect size (number of missing atoms).
			11,500~nm$^2$ of valid area was analysed before and 17,400~nm$^2$ after the irradiation.
			The inset shows a magnified view of the defect density for impurity atom substitutions with the lowest densities.
		}
		\label{fig:statistics}
	\end{figure}
	
	All recorded frames from pre- and post-irradiation were analysed using a CNN-assisted defect identification pipeline similar to the ones described in Refs.~\cite{ziatdinov_Deep_2017,trentino_AtomicLevel_2021,propst_Automated_2024,speckmann_Combined_2023,speckmann_Electron_2025,joudi_CorrugationDominated_2025} that relies on atomic number ($Z$)-contrast~\cite{krivanek_Atombyatom_2010} to differentiate between atoms and defects in the images (see Methods for more details).
	Supplementary Fig.~S3 illustrates the CNN data analysis workflow performed on every dataset in this study, and Supplementary Fig.~S4 showcases the defect-classification method.
	A comparison between the CNN-output defect density as a function of defect size, in non-irradiated and irradiated samples, is shown in Fig.~\ref{fig:statistics}a and Fig.~\ref{fig:statistics}b respectively.
	The individual datasets and their defect distributions acquired from every region in this study are also shown in Supplementary Fig.~S1.
	One particular region was exposed to the electron beam twice and is thus excluded from the post-irradiation statistics presented in Fig.~\ref{fig:statistics}b (see Discussion for more details).
	The as-prepared samples show a non-negligible defect density of 0.021~nm$^{-2}$ that could be explained, at least in part, by unavoidable electron-beam-induced defects along with the possibility of intrinsic defects being present in the material either from the growth or introduced during the sample preparation or cleaning procedure.
	The intrinsic defect distribution shows 55.6\% monoatomic defects, where 24.8\% are nitrogen single vacancies ($\mathrm{V_N}$), followed by 11.2\% boron single vacancies ($\mathrm{V_B}$), with an additional 9.4\% of silicon atom substitutions in boron sites ($\mathrm{Si_B}$) and 7.9\% and 2.3\% carbon atom substitutions, respectively in nitrogen and boron sites.
	The next most significant defect type is a complex vacancy structure with four or more missing atoms ($\mathrm{V_{4+}}$) at 21.5\%, followed by triple vacancies and double vacancies ($\mathrm{V_{BN}}$) accounting for 12.8\% and 8.4\%, respectively.

	\begin{table}[t]
		\centering
		\caption{\textbf{Defect statistics before and after irradiation.} Intrinsic values are written under `Before' and total post-irradiation values under `After'.}
		\label{tab:defect_stats}
		\begin{tabular}{
				|c|c|c|c|c|c|
			}\hline
			{\bf Size} &
			{\bf Type} &
			{\bf Before~(nm$^{-2}$)} &
			{\bf Before~(\%)} &
			{\bf After~(nm$^{-2}$)} &
			{\bf After~(\%)} \\
			\hline\hline
			\multirow{1}{*}{\textbf{1}}
			& $\mathrm{V_B}$  & 0.0024 & 11.21 & 0.0891 & 39.01 \\
			& $\mathrm{V_N}$  & 0.0053 & 24.77 & 0.0180 & 7.88 \\
			& $\mathrm{Si_B}$  & 0.0020 & 9.35 & 0.0354 & 15.50 \\
			& $\mathrm{Si_N}$  & 0 & 0 & 0.0003 & 0.13 \\
			& $\mathrm{C_B}$  & 0.0005 & 2.34 & 0.0003 & 0.13 \\
			& $\mathrm{C_N}$  & 0.0017 & 7.94 & 0.0002 & 0.09 \\\cline{2-6}
			& \textbf{Total} & \textbf{0.0119} & \textbf{55.61} & \textbf{0.1428} & \textbf{62.74} \\ 
			\hline
			\multirow{1}{*}{\textbf{2}}
			& \bm{$\mathrm{V_{BN}}$}     & \textbf{0.0018} & \textbf{8.41} & \textbf{0.0424} & \textbf{18.56} \\
			\hline
			\multirow{1}{*}{\textbf{3}}
			& $\mathrm{V_{BNN}}$ &  0.0023 &  10.75 &  0.0167 &  7.31 \\
			& $\mathrm{V_{BBN}}$ &  0.0006 &  2.80 &  0.0078 &  3.42 \\\cline{2-6}
			& \textbf{Total} & \textbf{0.0027} & \textbf{12.79} &\textbf{0.0245} & \textbf{10.73}  \\
			\hline
			\multirow{1}{*}{\bm{$\ge{4}$}}
			& \bm{$\mathrm{V_{4+}}$} & \textbf{0.0046} & \textbf{21.50} & \textbf{0.0180} & \textbf{7.87} \\
			\hline\hline
			\multicolumn{2}{|c|}{\textbf{TOTAL}} & \textbf{0.0214} & \textbf{100} & \textbf{0.2284} & \textbf{100} \\
			\hline
		\end{tabular}
	\end{table}

\begin{figure}[t]
		\centering
		\includegraphics[width=1\linewidth]{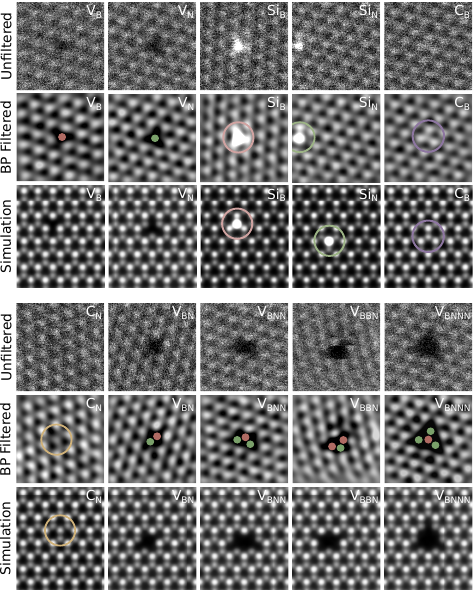}
		\caption{{\bf Example STEM-MAADF images and corresponding image simulations of typical defects acquired by the semi-automated imaging tool.}
			Two sets of images displaying raw images on the top, band-pass (BP) filtered (Fourier-space mask around the first-order diffraction spots) in the middle, and image simulations on the bottom.
			In the filtered images, the vacancy-type defects have been denoted with a solid coloured circle that represents the missing atoms (boron in red and nitrogen in green), and substitutional atoms have been circled.
			The raw and filtered images have been intensity-adjusted to portray the defects better.
			The multislice image simulations were made with the open-source \textit{ab}TEM package.
			All images have a FOV of $1.5\times 1.5$~nm$^2$.}
		\label{fig:defect_images}
	\end{figure}

	After irradiation, a total area of ca. 25,200~nm$^2$ was imaged with about 17,400~nm$^2$ being identified by the CNN as clean, monolayer hBN and the rest being masked out.
	A defect density of 0.228~nm$^{-2}$ was measured from the analysis resulting in an ion-induced defect density of 0.207~nm$^{-2}$, which is roughly ten times higher than the pre-irradiation value (0.021~nm$^{-2}$).
	The distribution of the different defect types and their densities after irradiation is shown in Fig.~\ref{fig:statistics}b.
	Now, 62.7\% of the defects are monoatomic comprising of 39.0\% boron single vacancies ($\mathrm{V_B}$), 15.5\% silicon atom substitutions in boron sites ($\mathrm{Si_B}$), 7.9\% nitrogen single vacancies ($\mathrm{V_N}$), and the remaining $<$1\% being silicon atom substitutions in nitrogen sites ($\mathrm{Si_N}$), and carbon atom substitutions in boron ($\mathrm{C_B}$) and nitrogen sites ($\mathrm{C_N}$). 
	Additional 18.6\% are double vacancies ($\mathrm{V_{BN}}$) and 10.7\% triple vacancies, with the remainder (7.9\%) being larger complex vacancies ($\mathrm{V_{4+}}$).
	We refer to Discussion and Methods for limitations on these results, particularly the identification of $\mathrm{C_B}$ and $\mathrm{C_N}$.
	A summary of all the different defect types and their contributions are tabulated in Table~\ref{tab:defect_stats}, showing the intrinsic values measured under `Before', and the total, non baseline-subtracted post-irradiation values under `After'.
	Every different type of defect that was identified in this study is shown in Fig.~\ref{fig:defect_images}, with the exception of the different complex vacancy structures ($\mathrm{V_{4+}}$). Raw, band-pass (BP) filtered and \textit{ab}TEM-simulated~\cite{madsen_AbTEM_2021} images are shown in two sets of blocks.
	
	\subsection*{Atomistic simulations}
	To estimate the expected defect densities, we conducted collision simulations of 150~eV Ar$^+$ ions with hBN using Ehrenfest molecular dynamics (see Methods and Supplementary Material for more details), implemented in the GPAW software package~\cite{mortensen_GPAW_2024,zobac_Ehrenfest_2025}.
	Contrary to our experimental results, but in agreement with other simulation studies~\cite{lehtinen_Production_2011,ghaderzadeh_Atomistic_2021}, we find no significant difference in the ejection of B or N atoms from the lattice for seven different impact points: centre of the B-N bond, halfway between the centre of the B-N bond and an atom (for both B and N), and four equally spaced points from the centre of the B-N bond to the centre of the hexagon (see Supplementary Fig.~S5).
	We also carried out DFT simulations to estimate whether there is a significant difference in the probability of filling $\mathrm{V_B}$ or $\mathrm{V_N}$ sites with C or O which could possibly explain the discrepancy in the experimentally measured and the theoretically predicted defect distribution.
    The simulations show that when a carbon adatom is placed near a vacancy, there is a significantly higher likelihood for it to fill a boron single vacancy as compared to a nitrogen single vacancy.
    For an oxygen adatom, there is no significant difference between the two.
    In the case of a double vacancy, a carbon adatom will always relax into the vacant boron site, resulting in an apparent nitrogen vacancy, whereas oxygen prefers the vacant nitrogen site, although not exclusively.
    As a summary, these simulations indicate that nitrogen single vacancies will be overestimated if there is a significant number of C or O impurity atoms incorporated in the created defects but not recognized during the analysis.
    Since we experimentally observe more boron single vacancies and also more impurity atom filling of boron sites, the simulations are unable to fully capture the defect creation dynamics.
	See Methods for more details regarding the simulation results.
    
	\section*{Discussion}

	Since electron irradiation during imaging is known to introduce defects into hBN~\cite{meyer_Selective_2009,jin_Fabrication_2009,kotakoski_Electron_2010}, even under ultra-high vacuum and at an acceleration voltage of 60~kV~\cite{bui_Creation_2023,cretu_Structure_2015,byrne_Fabrication_2026}, this effect needs to be correctly taken into account.
	To minimize the number of imaging-induced defects, each dataset acquired using the semi-automated imaging tool was taken from a region that had previously not been exposed to the electron beam.
	We estimate the beam current during the image acquisitions to be between 134 and 191~pA and thus expect 0.024 -- 0.034~nm$^{-2}$ single vacancies to be induced by the electron beam within the respective FOV during the imaging process (see Methods for a detailed calculation).
	This is comparable to the measured pre-irradiation defect density (0.021~nm$^{-2}$) in our as-prepared hBN samples.
	We note, however, that it is unlikely that all recorded pre-irradiation defects are introduced by the electron beam, mainly because larger defects (size $\ge{3}$) make up $\sim$40\% of all defects, and are unlikely to form during the acquisition of a single image~\cite{bui_Creation_2023, javed_Origin_2026a}.
	The total number of defects created in hBN during the irradiation can be larger than the above estimate as it only represents defects introduced within the FOV and not during stage movement between the recording of images.
	To check that this estimation is valid, we selected one region to be imaged pre- and post-irradiation.
	In that region, shown in Fig.~\ref{fig:methods_plasma}a and ~\ref{fig:methods_plasma}d (before and after irradiation, respectively), we measure a post-irradiation defect density that is higher by 0.04~nm$^{-2}$, compared to the total post-irradiation defect density elsewhere (see Supplementary Fig.~S1a,d. for the plots).
	This difference is two to three times larger than the estimate for electron-beam-induced defects, even when the dose on the sample during the stage movement is taken into account.
	A part of this could be explained by additional beam damage due to tuning during the set up of the automated imaging tool (see Methods), but another explanation could be that it is easier to damage a defective material than a pristine one, as reported in Ref.~\cite{kotakoski_Electron_2010}.
	
	Carbon and silicon atoms from the ubiquitous contamination, possibly mobilised during irradiation, are known to migrate on the surface of 2D materials and fill vacancies~\cite{inani_Silicon_2019a}, especially at elevated temperatures~\cite{postl_Indirect_2022}.
	Even though our UHV heating procedure removes surface contamination reliably, remnants are still present on defects, metal nanoclusters, grain boundaries, and multilayer regions even at temperatures as high as 550$^{\circ}$C.
	Achieving cleaner samples beyond this, is limited by the defects and metal contamination that is most likely introduced during the growth of the material~\cite{irschik_Atomically_2026}.
	Since we do not see nearly as many substitutional atoms in the pre-irradiation datasets (0.004~nm$^{-2}$ compared to 0.036~nm$^{-2}$), we believe this process occurs during and/or immediately after the irradiation but before the STEM imaging.
	We find that a significant number of brighter atom substitutions are particularly prominent in $\mathrm{V_B}$ sites, most of them assumed to be Si atoms from their ADF contrast~\cite{ahmadpourmonazam_Substitutional_2019}.
	As shown in the defect statistics, we only found 5 $\mathrm{V_N}$ sites being filled by Si ($\mathrm{Si_N}$), compared to 616 $\mathrm{V_B}$ sites ($\mathrm{Si_B}$).
	Moreover, we observe that more $\mathrm{Si_B}$ defects are found closer to contamination regions than away from them.
	Fig.~\ref{fig:scanmap}b is an image supporting this observation, showing that regions closer to contamination patches (top right and bottom middle) contain many vacancies that have been filled by brighter atoms while the regions further away from contamination (top left and middle) contain vacancies as is.
	This could indicate that the impurity atoms originate from the contamination, and are mobilised by the ions which then diffuse and fill vacancies that are closest to them.
	Unlike Si, it is much more difficult to detect carbon-filled defects ($\mathrm{C_B}$ and $\mathrm{C_N}$) through the CNN-assisted analysis, due to the similar atomic number of B, C and N (and therefore ADF contrast~\cite{krivanek_Atombyatom_2010}).
	The post-irradiation datasets were also generally more noisy and had higher intensity fluctuations in the images, which added to the difficulty in detection of these carbon substitutions (see Methods for more details).
	Nevertheless, we were able to detect 20 $\mathrm{C_N}$ and 6 $\mathrm{C_B}$ in the pre-irradiation datasets, and 4 $\mathrm{C_N}$ and 6 $\mathrm{C_B}$ in the post-irradiation datasets.
	
	We note that the identification of these defects, via the CNN and its post-processing steps, has a slight bias towards the smaller defects.
	This is due to manually setting threshold values for each image to maximise the accuracy of identifying defects with one or two missing atoms, which can potentially artificially increase the number of missing atoms detected in larger vacancies in that same frame.
	We stress that this over-counting, primarily in the 4+ category, only artificially increases the size of those vacancies but does not introduce new defect counts in the statistics.
	Since images acquired from ion-irradiated samples are often more difficult to properly analyse due to large intensity variations in the images and within a particular dataset (see Supplementary Fig.~S6 and~S7 for examples), this effect is limited to post-irradiation results.
	
	From the geometry of the hBN lattice and the expected size of ion-induced defects of one or two atoms, if a subsequent ion was to impinge on hBN it would have to be within two unit cells (twice the lattice constant, i.e., 0.5~nm) of an existing defect in order to possibly enlarge that defect.
	If it impacted further away, it would encounter pristine hBN possibly creating a new defect.
	At the post-irradiation defect density, the average inter-defect distance is roughly 2.4~nm and each defect occupies an average area of 4.4~nm$^2$. Dividing the area within which an ion must impinge in order to encounter defective hBN by the average area per defect, we get a probability of 18\% that a subsequent ion would encounter defective hBN.
	The rest of the time, it would encounter pristine hBN.
	We note that this has implications for defect size three and above as roughly 19\% of defects after irradiation belong to this category, and hence some fraction of these defects could be explained due to this effect where a subsequent ion might encounter a single or double vacancy and increase its size.
	However, if the existing defects are charged~\cite{huang_Defect_2012a}, their influence on the defect creation process can reach further away than the two unit cells due to their influence on the local bonding environment as well as charge transfer between the incoming ion and the lattice. Hence, there may be changes to the displacement threshold energy even when the ion impact leads to the creation of a new defect.
	
	In many images, the intensities of some nitrogen sites surrounding boron single vacancies appear comparable to or even higher than those of nitrogen sites in the pristine lattice (see Supplementary Fig.~S8).
	This is unexpected, as atoms adjacent to a vacancy should exhibit slightly lower intensities due to the missing probe-tail scattering contribution from the absent neighbouring atom~\cite{krivanek_Aberrationcorrected_2015}, and increased vibrational smearing of the under-coordinated edge atoms.
	This observation raises the possibility that some boron vacancies could be surrounded by slightly heavier atoms, most likely oxygen~\cite{hofer_Direct_2019}, consistent with the findings reported by Li et al.~\cite{li_Prolonged_2023}, where 10 out of 12 of their measured boron single vacancies had oxygen atom substitutions in nitrogen sites around them.
	Recently, ab initio calculations using spin-polarised DFT have also predicted that the oxygen atom substitution in a nitrogen site ($\mathrm{O_N}$) could be responsible for the 3.5~eV photoluminescence peak in hBN~\cite{maciaszek_Substitutional_2026}.
	To qualitatively demonstrate this observation, we refer to Supplementary Fig.~S9, which shows \textit{ab}TEM image simulations of a boron single vacancy surrounded by nitrogen or oxygen atom substitutions under different imaging conditions (idealized probe, probe tail contributions, vibrational smearing, realistic conditions).
	As noted above, both probe tail contributions and vibrational smearing result in lower intensities for edge nitrogen atoms compared to lattice nitrogen atoms; however, oxygen atom substitutions result in similar or higher intensities than lattice nitrogen atoms, similar to what we show in Supplementary Fig.~S8.
	However, we cannot unambiguously discern the atomic species of these higher intensity atoms from the experimental data, as our analysis methods were optimized for unbiased vacancy detection rather than for discerning subtle intensity differences, and we must leave this for a future study.

	We find defect distributions similar to other studies, although the intrinsic and induced defect densities vary significantly in literature.
	Wang et al.~\cite{wang_Photoluminescence_2018} report a surprisingly high intrinsic defect density of 0.14~nm$^{-2}$ where $\mathrm{V_B}$ accounts for 78.7\% of all the defects measured.
	Since they do not take electron-beam damage at 60~kV (boron single vacancies are about 1.4 times as likely to be created than nitrogen single vacancies)~\cite{bui_Creation_2023}, and chemical effects (residual oxygen in their microscope column)~\cite{javed_Origin_2026a} into account, this could possibly explain the number of vacancies and the prevalence for B site vacancies.
	Liang et al.~\cite{liang_SiteSelective_2025} irradiate multilayer hBN with 500~keV He$^+$ ions where the interaction is dominated by inelastic effects (electronic stopping).
	They report a defect density of ca. 0.09~nm$^{-2}$ for a fluence of 30 ions nm$^{-2}$ which would correspond to a defect creation probability of around 0.3\% per ion.
	Surprisingly, their induced defect distribution is similar to the one in our study; also showing that B sites are more prone to vacancy formation which are then also more prone to adatom absorption/interacalants, relative to N sites.
	While their post-irradiation defect density in multilayer hBN is lower than the intrinsic defect density in monolayer samples reported by Wang et al.~\cite{wang_Photoluminescence_2018}, they state that the defects in their irradiated samples are of a much more complex structure and that impurity-type defects are commonly observed in their irradiated samples.
	Byrne et al.~\cite{byrne_Atomic_2025} irradiated monolayer hBN with He$^+$ and Ne$^+$ at 25~keV with different doses and report defect densities but no defect-type distribution.
	Extracting the ion-induced defect density from the data they report, results in a ca. 0.17\% probability of creating a defect per ion.
	
	Our experimental results are, however, in clear contradiction with theoretical studies based on analytical-potential molecular-dynamics simulations, where it was found, depending on the potential used, that an impinging Ar ion ($\approx$ 100 eV) has a 75--80\% chance to create a defect with the probability for creating a boron or nitrogen single vacancy being 17--20\% each and about 40\% for a double vacancy~\cite{lehtinen_Production_2011}.
	Another study used 150~eV Ar irradiation and predicted it should create roughly twice as many N single vacancies than B single vacancies or double vacancies~\cite{ghaderzadeh_Atomistic_2021}.
	With an ion fluence of 0.8$\pm$0.2~nm$^{-2}$ and an ion-induced defect density of 0.207~nm$^{-2}$, we show a defect-creation probability of about 26$\pm$5\%. Ignoring vacancy filling, the resulting defect distribution shows a ca. 6:1 ratio of B:N single vacancies and a roughly 3:1 ratio of single to double vacancies.
	The uncertainty for the defect creation efficiency is relatively large due to the measured ion-beam current being highly dependent on the exact position of the Faraday cup, which is prone to misalignment.
	Additionally, the position of the sample during the irradiation can vary, contributing to further uncertainty.
	There may also be cases where ions are neutralized on the way to the cup/sample but retain their kinetic energy and could thus induce defects without contributing to the current.
	Although, we don't expect these to have a major contribution to the current and to the defect creation efficiency estimate.
	We note that in Ref.~\cite{lamprecht_Uncovering_2025} we observe a similar defect creation efficiency using ultra-low-energy ($\approx$170~eV) He$^+$ on MoS$_{\mathrm{2}}$ using the same set-up, although the irradiation effect is not that straightforward to approximate in MoS$_{\mathrm{2}}$ due to cascading effects.
	
	In comparison to our previous work where graphene was defect-engineered using Ar$^+$ ions at a similar energy and with the same setup~\cite{trentino_AtomicLevel_2021,joudi_CorrugationDominated_2025}, we find that the defect-size distributions are curiously different.
	In the case of graphene, we primarily create double vacancies (ca. 3.5 times more than single vacancies).
	At the energy regime that we are irradiating with, we are comfortably above the displacement threshold energy of carbon atoms in graphene and also of boron and nitrogen atoms in hBN.
	Assuming the same knock-on type of irradiation damage (and displacement cross-sections) for both materials, one would expect a similar defect size to be created in both cases.
	However, we find the opposite in hBN where we have ca. 3.4 times more monoatomic defects than double vacancies.
	One explanation could be that, in the case of hBN, we might be imaging a `smaller' defect than what was created by the ion irradiation.
	If we assume that double vacancies are created in hBN upon irradiation, just like in graphene, then vacancy filling by carbon, oxygen, and silicon atoms, as discussed earlier, could fill one (or two) of the atomic sites in a double vacancy and form a monoatomic defect that is then imaged.
	This would, of course, also be occurring for the larger defect sizes, resulting in the size distribution shifting towards smaller defects.
	However, as explained in our previous work~\cite{trentino_AtomicLevel_2021}, bond rearrangement in graphene allow vacancies to migrate and agglomerate, forming larger defects.
	Some part of the differences in the defect sizes between graphene and hBN could also be explained due to this vacancy agglomeration, since bond rearrangement is not possible in hBN~\cite{jin_Fabrication_2009}.
	
	Finally, it is typically assumed that singly-charged ion impacts can be modeled without taking charges into account, because prior to impact, the ion receives an electron from the sample, hence neutralizing it~\cite{lehtinen_Effects_2010}.
	With DFT-based molecular dynamics, the calculated displacement threshold energy in a neutral hBN lattice is lower for boron than for nitrogen (19.36~eV vs. 23.06~eV, respectively)~\cite{kotakoski_Electron_2010}.
	Since energy transfer from Ar to N is more efficient than to B, due to momentum conservation and the higher mass of N (14.007 amu vs 10.811 amu), this fact by itself is not sufficient to explain our experimental results.
	Even when the charge of the impinging ion was included in Ehrenfest simulations described above, the results showed approximately equal probability for displacing either boron or nitrogen atoms from hBN following an impact by a 150~eV Ar$^+$ ion.
	However, while in conductive samples, the missing charge due to the neutralization of the ion is replenished quickly through the material, this is not necessarily true for an insulator such as hBN, where the charge may remain localized in the vicinity of the impact point, similar to what has been suggested for MoS$_2$~\cite{kretschmer_Formation_2020}.
	While the average inter-defect distance in our samples is sufficiently far to allow individual defects both to be created by the ions as well as to be detected as such through atomic-resolution imaging, the corresponding density (1 defect per ca. 4.4~nm$^2$) of possibly charged defects~\cite{huang_Defect_2012a} is likely to alter the bonding environment, causing a change in the displacement threshold energy.
	Similarly, it may have an influence on the neutralization of the approaching charged ion, raising further complications to an adequate computational description of the defect-creation process.
	However, it is also possible that the mean-field description of Ehrenfest simulations is insufficient to capture the bond breaking process in hBN under ion irradiation, and that the pathways associated with different quantum states during the charge transfer processes would need to be modeled separately for an accurate description.

	To conclude, we demonstrated that ultra-low-energy Ar$^+$ irradiation is a suitable method for defect engineering hBN with a high selectivity for boron single vacancies.
	The pre-irradiation defect density in the as-prepared and cleaned samples was found to be 0.021~nm$^{-2}$ which also includes electron-beam damage, estimated to induce 0.024 -- 0.034~nm$^{-2}$ single vacancies within the FOV during semi-automated imaging of the samples.
	The ion irradiation fluence was chosen to be high enough to separate the irradiation-induced defects from the intrinsic ones, but sufficiently low to mostly avoid ions impinging on already defective sites.
	After irradiation, we measured a defect density of 0.228~nm$^{-2}$ of primarily monoatomic defects with a preference for $\mathrm{V_B}$.
	Our results challenge earlier computational work predicting either similar formation probabilities for $\mathrm{V_N}$ and $\mathrm{V_B}$ and a higher probability for $\mathrm{V_{BN}}$~\cite{lehtinen_Production_2011}, or a predominance for $\mathrm{V_N}$ over $\mathrm{V_B}$ and $\mathrm{V_{BN}}$ under Ar irradiation at similar energies~\cite{ghaderzadeh_Atomistic_2021}.
	Surprisingly, the predominance of $\mathrm{V_B}$ was not only experimentally shown here, but has also been reported for much higher ion energies~\cite{liang_SiteSelective_2025} and in the case of intrinsic defects~\cite{wang_Photoluminescence_2018}.
	Our results also reveal a largely under-studied phenomenon of vacancy filling by impurity atoms (silicon, carbon, and possibly oxygen) as a result of ion irradiation and show that more research is needed both experimentally and computationally to fully understand the mechanisms involved in the defect-engineering of hBN.
	We also show that the irradiation results in hBN are much more complex than in graphene~\cite{trentino_AtomicLevel_2021,joudi_CorrugationDominated_2025}, where double vacancies dominate and vacancy filling by impurity atoms does not play as big of a role.
	Above all, we demonstrate that ultra-low-energy ion irradiation creates defects far more efficiently than higher-energy irradiation experiments, and that careful interpretation is needed when using electron-beam characterisation, as the measurement process itself can induce defects~\cite{meyer_Selective_2009,jin_Fabrication_2009,kotakoski_Electron_2010,bui_Creation_2023}.
	
	\section*{Methods}
	All experimental steps described here were conducted for two samples.
	\subsection*{Sample preparation and cleaning}
	Commercially available CVD-grown monolayer hBN on copper (Cu) foil (purchased from Sigma-Aldrich) was used as the source material.
	The transfer was performed via the electrochemical delamination method~\cite{wang_Electrochemical_2011,irschik_Atomically_2026}, where the foil was first cut into ca. 3$\times$3~mm$^2$ squares to match the size of standard TEM grids.
	These were spin-coated with two drops (ca. 20~$\mu$L) of polymethyl methacrylate (PMMA), which acts as a supportive film on top of hBN to prevent folding and damage during the transfer.
	After curing on a hot plate for 1~h at 150$^{\circ}$C, less than 1~mm was cut off from all edges of the squares to remove any PMMA spillover.
	A 1M NaOH solution was used as the electrolyte with a platinum (Pt) wire as the anode and the Cu foil (PMMA/hBN/Cu stack) as the cathode.
	Using 4~V, the generation of H$_2$ gas bubbles at the Cu foil gently delaminates the PMMA/hBN stack.
	The delaminated stack was transferred to a bath of deionized (DI) water to wash off solvent residue and was then transferred onto Quantifoil (QF) gold (Au) TEM grids that have an amorphous carbon supportive membrane with 1.2~$\mu$m holes. 
	After drying the remaining solvent on a hot plate (150$^{\circ}$C), the PMMA was dissolved by immersing the sample in a hot acetone bath (50$^{\circ}$C) for 1~h, followed by a room-temperature isopropyl alcohol (IPA) bath for 30 min.
	
	Upon insertion into the vacuum system used for all experiments~\cite{mangler_Materials_2022}, the samples were baked for ca. 10~h at 150$^\circ$C in vacuum to remove water and loosely bound contamination.
	Prior to imaging, the samples were baked again, but now in a UHV heater~\cite{irschik_Atomically_2026} at 250$^{\circ}$C for 180~min to achieve clean monolayer regions required for this study.
	This bake is essential for removing more strongly bound surface contamination and uncovering more of the underlying monolayer hBN that will be examined using STEM.

	\subsection*{Ion irradiation}
	The samples were irradiated with an ultra-low-energy Ar$^+$ beam using a SPECS ECR-HO microwave plasma source operated in hybrid mode with an anode voltage of 0~V and extractor voltage of -60~V.
	A schematic of the set-up is shown in Fig.~\ref{fig:methods_plasma}b.
	The irradiation is carried out at room temperature.
	Directly comparing the irradiation parameters shown in this work to the ones in Ref.~\cite{langle_Defectengineering_2024,propst_Automated_2024} is challenging as the plasma source needed to be maintained and modified between the experiments.
	Nevertheless, for the experiments reported here, the pressure in the chamber was kept between 3.6 and 4.2 $\times$$10^{-6}$~mbar.
	In all cases, the plasma was stabilized for at least 20 min, the beam was measured, and the sample was inserted into the ion beam path for 150~s, after which the beam was measured again.
	
	To estimate the ion energies, the current was measured as a function of a bias voltage applied to a Faraday cup (see Fig.~\ref{fig:methods_plasma}c).
	We assume the plasma leaving the source to consist of an ion beam with a Gaussian energy profile as well as electrons.
	With the applied positive bias, electrons are accelerated towards the Faraday cup, past a repelling magnet (used to filter electrons), which leads to a linear increase of negative charge carriers.
	When the decelerating bias matches the kinetic energy of the ions this appears as a drop in the measured current as the positively charged ions no longer reach the cup.
	At each bias, the current at the cup corresponds to the sum of all impinging charge carriers (electrons and ions).
	Thus the measured curve is fitted with the integral of the sum of a linear function (for the electron contribution) and a Gaussian (for the ion contribution), corresponding to a second-order polynomial and an error function.
	The amplitude of the error function is the ion fluence at the position of the Faraday cup.
	
	The Faraday cup is placed in the same lateral position as the sample with a slightly larger distance from the source.
	According to the device manual, the spot diameter $D$ (in mm), is given by
	\begin{equation*}
		D = 2d\,\tan(\alpha) + 25,
	\end{equation*}
	where $d$ is the distance from the source (in mm) and $\alpha$ is the opening angle dependent on the energy, which is 60$^{\circ}$ at 150~eV.
	The distance from the source to the sample is 20$\pm$2~mm and to the Faraday cup 25$\pm$2~mm, which leads to a difference in the measured and the actual current at the sample of a factor of ca.~1.4$\pm$0.3 (uncertainties arise from the measurements of the distance with a standard ruler).
	Multiplying the amplitudes (17.28 and 21.89)~nA by this factor and the irradiation time (150~s) and dividing it by the area of the opening aperture (diameter = 6.35$\pm$0.02~mm) to the Faraday cup, we get the number of charge carriers per area, 0.7$\pm$0.1 and 0.9$\pm$0.2~nm$^{-2}$ (uncertainties arise from the measurement of the diameter with a caliper).
	The average ion fluence is then calculated to be 0.8$\pm$0.2~nm$^{-2}$.
	All beam currents (raw data and fits) are shown in Supplementary Fig.~S2.
	At the position of the sample, the ion plasma cone diameter is estimated to be 9.4~cm, while the TEM grid has a diameter of 3~mm.
	We thus assume an approximately constant flux throughout the entire sample.

	\subsection*{Scanning transmission electron microscopy}
	All electron microscopy images were recorded with the Nion UltraSTEM 100 in Vienna at 60~kV in ultra-high vacuum (10$^{\mathrm{-10}}$~mbar) using a medium-angle annular dark-field (MAADF) detector with a semi-angular range of 60--200~mrad.
	The convergence semi-angle was 35~mrad.
	The beam current was not recorded during the day of the imaging, but it was measured on three different days around the time of recording the data.
	At least five subsequent data sets recorded on the direct-electron detector Dectris ARINA were used to estimate the beam current~\cite{susi_Quantifying_2025} on those three days (191 $\pm$ 5 pA, 134 $\pm$ 3 pA, and 145 $\pm$ 6pA).
	
	For obtaining defect densities and distributions, we used a semi-automatic image acquisition tool~\cite{mittelberger_Automated_2017}.
	With this method, the algorithm records images in a predefined area of the sample along a serpentine path illustrated in Fig.~\ref{fig:scanmap}a.
	We defined the grid of recorded frames to avoid overlap between them by leaving an offset of 2.5 frames (ca. 6 nm) between the captured images.
	This is particularly important due to the hysteresis of the stage when changing direction~\cite{mittelberger_Automated_2017}.
	During the image acquisition, minor adjustments of the fine focus and the correction of astigmatism and coma were done manually, however always after acquiring a frame, but before moving to the new FOV.
	A sleep time of 3~s between frame acquisitions was set to avoid stage drift and artifacts in the images when moving from one imaging position to the next.
	Before the acquisition starts, the beam rests in the middle of the FOV.
	Because of this, the center of each subsequent frame is exposed to the electron beam for 1--2~s before the scan is started contributing to potential electron-beam-induced defects.	
	All images were recorded with a pixel dwell time of 16~$\mathrm{\mu s}$ and 512$\times$ 512 pixels.
	Adding the flyback time (240~$\mu$s) at the end of each of the 512 scan lines and the sleep time (taken into account as 2$\pm$1~s of beam exposure within the imaged  FOV) we calculate a total exposure time of 6.3$\pm$1.0~s per frame.

	The images were recorded with a  field of view of 4$\times$4~$\mathrm{nm}^2$, which was confirmed using Fourier-transform calibration~\cite{madsen_Fourierscalecalibration_2022}.
	For the highest beam current of $191 \pm 5$~pA, this results in an electron dose of
	(4.7 $\pm$ 1.1) $\times$ 10$^6$~electrons \AA$^{{-2}}$ within the field of view (the uncertainty is dominated by the 1~s uncertainty chosen for the speed of the stage movement).
	Using the displacement cross section from Ref.~\cite{bui_Creation_2023}, we estimate that $191 \pm 5$~pA, corresponding to $4.7\pm 1.1 \times10^{6}$ electrons per \AA$^2$, should create	ca. 0.034 $\pm$ 0.008 single vacancies per nm$^2$, with 41\% $\mathrm{V_N}$ and 59\% $\mathrm{V_B}$. 
	Respectively, for
	$134 \pm 3$~pA, 
	corresponding to ca. $3.3\pm 0.8\times10^{6}$ electrons per \AA$^2$, $0.024\pm 0.006$ defects per nm$^2$ are induced,
	and for
	$145 \pm 6$~pA, 
	corresponding to $3.6\pm 0.9 \times10^{6}$ electrons per \AA$^2$,  $0.026 \pm 0.006$ defects per nm$^2$ are induced, all with the same vacancy ratio as mentioned above.
	This is, of course, only true for hBN without defects, as the cross section increases significantly if the atoms are at an edge~\cite{javed_Origin_2026a}.
	The displacement threshold energy of boron and nitrogen atoms at monovacancy edges are reduced (12.92~eV and 15.00~eV, respectively) compared to that in pristine regions (19.36~eV and 23.06~eV, respectively)~\cite{kotakoski_Electron_2010}.
	The same decreasing trend in the displacement threshold energy of edge atoms is seen as the size of the vacancy is increased.
	Therefore, to reduce electron-beam-induced damage we would ideally need to blank the beam during the time between images and, if possible, also during the flyback time.

	\subsection*{Semi-autonomous image analysis}
	Supplementary Fig.~S3 shows the entire pre-processing workflow described below up until the defect classification via thresholds.
	The raw data collected from the semi-automated imaging tool was manually inspected and images of poor quality and images containing $>$50\% contamination were discarded.
	The images were then pre-processed to mask out contamination by setting a global threshold.
	An image of pristine hBN with no contamination was chosen and a double-Gaussian filter ($\sigma_1$ = 0.25, $\sigma_2$ = 0.15, and weight = 0.3) \cite{krivanek_Atombyatom_2010} was applied to it.
	All intensities higher than the set global threshold on the histogram were masked out and classified as contamination. 
	Before passing the images to the convolutional neural network (CNN), in order to identify the lattice, vacancies, and substitutional atoms more easily, all images were also filtered with the same double-Gaussian filter.
	The analysis was mainly conducted in the way as described in Ref.~\cite{propst_Automated_2024,speckmann_Combined_2023,joudi_CorrugationDominated_2025} using a CNN similar to the one used in Ref.~\cite{trentino_AtomicLevel_2021}.
	We refer the reader to Ref.~\cite{propst_Automated_2024} for a detailed explanation of the CNN and its lattice-identification process.
	For hBN, the used CNN does not directly classify defects in the lattice but outputs atomic positions for a perfect lattice, see Supplementary Fig.~S6 for examples of the correct lattice recognition (panels (a) to (e)) as well as a few instances where it fails (panels (f) to (j)).
	Images where the lattice recognition failed were discarded.
	Based on the intensity at the lattice sites ($Z$-contrast~\cite{krivanek_Atombyatom_2010}), three thresholds ($\mathrm{T_1}$, $\mathrm{T_2}$ and $\mathrm{T_3}$) are used for the classification of atoms, vacancies, and substitutional atoms.
	In each sublattice, the rules for defect quantification are:
	$\mathrm{V_B} < \mathrm{T_1}$, $\mathrm{T_2} <\mathrm{C_B} < \mathrm{T_3}$, $\mathrm{T_3} < \mathrm{Si_B}$ for the boron sublattice and
	$\mathrm{V_N} < \mathrm{T_1}$, $\mathrm{T_1} <\mathrm{C_N} < \mathrm{T_2}$, $\mathrm{T_3} <\mathrm{Si_B}$ for the nitrogen sublattice.
	The thresholds (columns V) and the different steps can be seen at work in Supplementary Fig.~S4 and Fig.~S7.
	At first, global thresholds are set, however, manual intervention and adjusting is needed, particularly in the post-irradiation datasets as we observed a large variation in intensity of the individual images within a particular dataset.
	In some instances misidentification could not be avoided, as sometimes the identified defects and heteroatoms do not agree with the image contrast at those positions.
	These few cases were still included in the statistics, some examples shown in Supplementary Fig.~S7. 
	This introduces the potential of human error and miscounting; however, the data was independently analysed four times with only minor differences in the statistics.
	The outputted positions of all the classified defects are saved and grouped together depending on their proximity from each other (per frame).
	If atoms were within a radius of 1.5 times the bond length, they were grouped together into a larger defect.
	With this, statistics on defect types and their sizes were made.
	We acknowledge that the defect densities of defect sizes `3' and `4+' have been slightly over-estimated by ca. 10--15\% in order to accommodate the identification of single and double vacancies, as explained earlier.

	\subsection*{STEM image simulations}
	ADF-STEM image simulations were performed based on independent atom potentials of pristine and defective hBN using the \textit{ab}TEM package~\cite{madsen_AbTEM_2021}. In all cases parameters were set to replicate common imaging conditions of Nion probe-corrected microscopes with an electron beam energy of 60 keV, a probe convergence semi-angle of 35 mrad, an ADF semi-angular range of 60--200~mrad, a source size of 0.03~nm, spherical aberration (Cs) of $\SI{3}{\micro\meter}$ and a dose of $10^6$~$\mathrm{e}^-$/\AA$^2$. To account for thermal diffuse scattering, we used the frozen-phonon model with 20 snapshots per image using a standard deviation of atomic displacements of 0.1~\AA. To account for partial temporal coherence of the non-monochromated beam we averaged over 12 Gaussian-distributed defocus values with a focal spread of 5 nm.

	\subsection*{DFT simulations}
	To identify the most favorable adsorption position for a carbon (C) adatom on an hBN layer in the vicinity of a nitrogen (N) or a boron (B) single vacancy, we systematically positioned the C atom around the respective vacancy sites.
	To simulate vacancies in hBN we built an 8 $\times$ 8 supercell with periodic boundary conditions, with $\Gamma$-point sampling in the reciprocal space.
	The explored adsorption region on the hBN layer formed a rectangular area of 0.50 $\times$ 0.72 nm$^2$ with its origin in the vacancy site, sufficiently covering the symmetric region around it.
	The C atom was placed at each point of a rectangular grid consisting of 11 points along the $x$ and $y$ directions at a height of 2~\AA~above the lattice.
	Each of the 121 resulting structures per defect type, corresponding to a unique grid point, was subsequently relaxed.
	Relaxation calculations were performed using density functional theory (DFT) within the projector augmented-wave (PAW) method as implemented in the GPAW software~\cite{mortensen_GPAW_2024}.
	The force tolerance was set to 0.05 eV/\AA, and the density tolerance to 10$^{-6}$.
	A double-zeta polarized basis set of linear combinations of atomic orbitals (LCAO) was employed, along with a grid spacing of 0.2~\AA~and the Perdew-Burke-Ernzerhof (PBE) exchange-correlation functional~\cite{perdew_Generalized_1996}.
	The C atom relaxed into the N vacancy in 8 cases and into the B vacancy in 31 cases.
	These results suggest that the C atom preferentially migrates to the B vacancy site over the N vacancy.
	
	To simulate the collision of Ar$^+$ ions with hBN, we employed Ehrenfest molecular dynamics as implemented in the GPAW software package~\cite{mortensen_GPAW_2024,zobac_Ehrenfest_2025}.
	A 6$\times$6 hBN supercell was used to model the target, with periodic boundary conditions applied laterally.
	The simulation cell was extended in the \textit{z}-direction to a total length of 100 Å to accommodate the ion trajectory and minimize interactions with periodic images.
	A double-zeta polarized basis set of linear combinations of atomic orbitals (LCAO) was used for all atoms, and a real-space grid spacing of 0.2 Å was applied.
	The Brillouin zone was sampled at the $\Gamma$ point.
	Impact sites were chosen to sample the irreducible area of the unit cell and are illustrated in Supplementary Fig.~S5 with the differently coloured circles (black, red, pink, blue, yellow, brown and green).
	The initial energy of the incident Ar$^+$ ion was set to 150~eV.
	The ionic charge state was initialized using constrained DFT~\cite{melander_Implementation_2016} to prepare Ar$^+$ as the initial state of the molecular-dynamics simulation.
	The simulations were spin-polarized, and the PBE exchange-correlation functional was employed.
	The time evolution was integrated using a time step of 40 as, and the simulations were continued until the projectile had either scattered back or transmitted through the layer.
	No thermostat was used for the simulations.
	In case of the impact points marked in yellow, brown, green, blue, and pink we find that atoms 5 and 6 get completely displaced from the lattice.
	Impact points marked in red and black cause fluctuations in the positions of atoms 5 and 6 but do not displace them.
	
	\section*{Data availability}
	The data supporting the findings of this study are made openly available at the following URL: https://phaidra.univie.ac.at/o:2343169
	
	\section*{Acknowledgments}
		This research was funded in part by the Austrian Science Fund (FWF) [10.55776/COE5 and 10.55776/P36264].
		ML acknowledges funding from the European Union’s Horizon Europe research and innovation programme under the Marie Skłodowska-Curie grant agreement No. 101210084.
		For open-access purposes, the author has applied a CC-BY public copyright license to any author-accepted manuscript version arising from this submission.
		
	\section*{Author contributions}
	ML and JK conceived the study.
	SC, ML, BMM and AB conducted the experimental work.
	SC, ML, BMM, JM, DP and JK performed the data analysis.
	JM and DP trained the neural networks used in this study.
	SC, CK, and PI prepared the samples.
	SC, ML, VH and CM developed the irradiation set-up.
	PI  developed the heating stage used for sample cleaning.
	FK developed a laser system used for sample cleaning in an earlier iteration of this project~\cite{langle_Defectengineering_2024}.
	VZ performed the \textit{ab initio} calculations.
	DL performed the \textit{ab}TEM image simulations.
	SC, ML, and JK wrote the manuscript, with contributions from all authors.
	LF, TS and JK supervised the study.
	
	\section*{Competing interests}
	The authors declare no competing financial or non-financial interests.

\def\bibsection{\section*{\refname}} 	
\bibliography{scibib}
\bibliographystyle{ieeetr}

\onecolumngrid
\clearpage
\setcounter{figure}{0}
\makeatletter 
\renewcommand{\thefigure}{S\@arabic\c@figure}
\makeatother




\maketitle
\onecolumngrid
\clearpage
\renewcommand{\thefigure}{S\arabic{figure}}
\setcounter{figure}{0}  

\clearpage
\section*{SUPPLEMENTARY MATERIAL \\ Image acquisition areas and defect statistics}

\begin{figure*}[b!]
	\centering
	\includegraphics[width=0.9\textwidth]{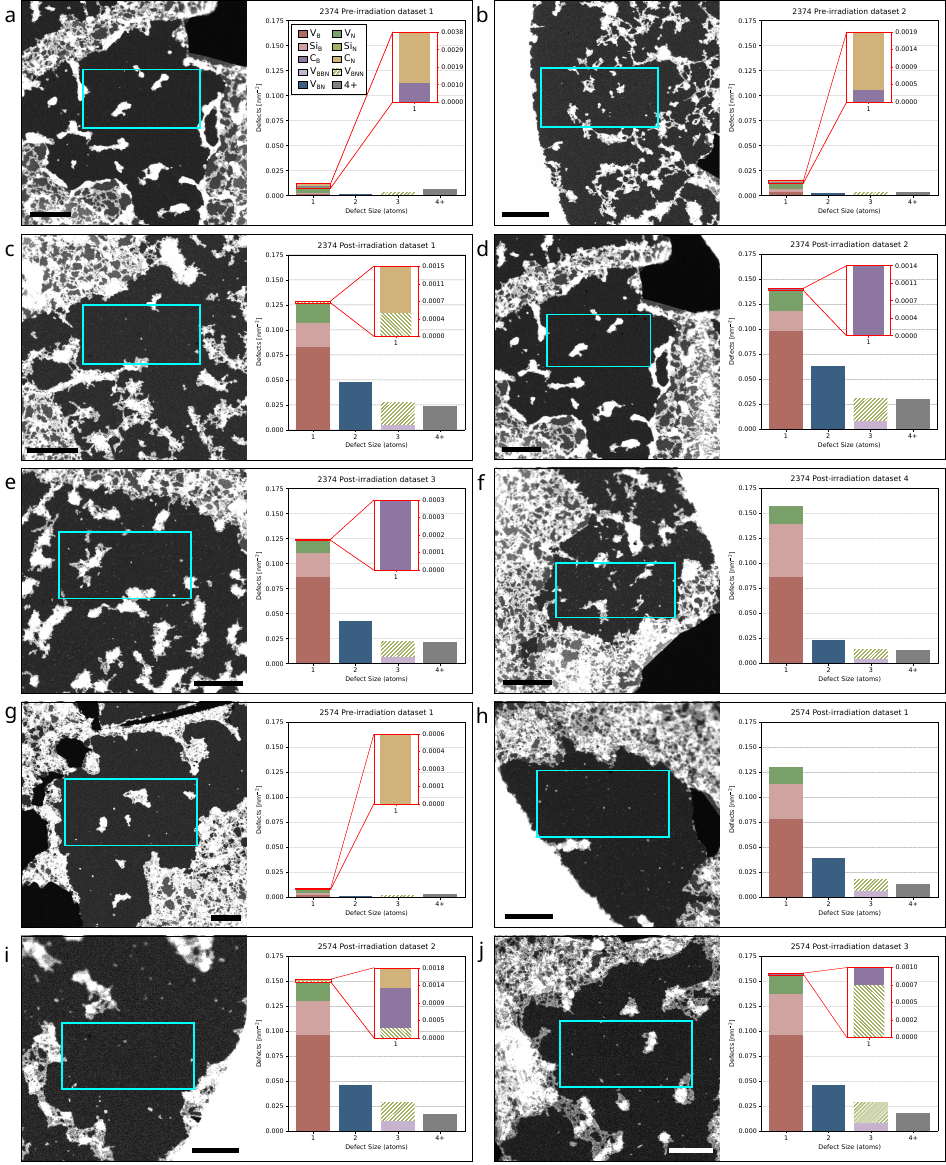}
	\caption{{\bf Areas used for semi-automatic imaging (marked with a cyan rectangle) alongside the defect-density statistics.}
		The area shown in panel (a) and (d) was exposed to the electron beam twice: once before irradiation and once again after irradiation. All other areas were only imaged and exposed to the electron beam once.
		The scale bar in all images corresponds to 50~nm.
	}
	\label{fig:overview_figure}
\end{figure*}

\clearpage
\section*{Beam Profiles}

\begin{figure*}[h!]
	\centering
	\includegraphics{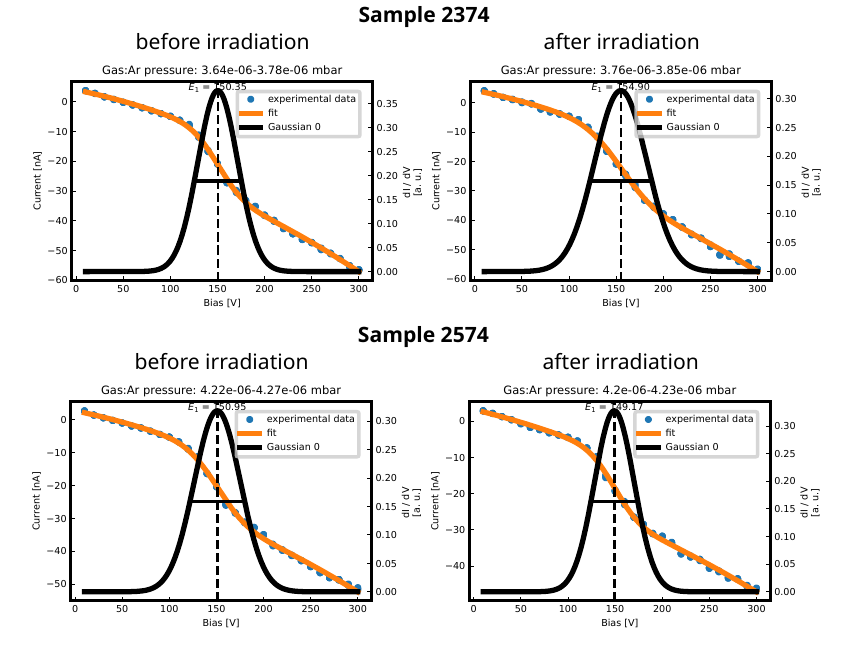}
	\caption{{\bf Beam-profile measurement raw data and fits.}
		The beam current (blue dots) and fit (orange line) as well as the Gaussian obtained by the fit to $\mathrm{dI/dV}$ (black line, right axis) are shown from left to right for the sample 2374, before and after irradiation, and sample 2574, before and after irradiation.
		The fit parameters are given in Table~\ref{tab:fits}.
	}
	\label{fig:beamshape_supplement}
\end{figure*}

\begin{table*}[h!]
	\centering
	\caption{\textbf{Fit parameters} of the beam-profile measurements before and after the irradiation of samples 2374 and 2574, corresponding to the plots shown in Fig.~\ref{fig:beamshape_supplement}.
	The mean of the distribution is denoted by $\mu$ and standard deviation by $\sigma$.}
	\begin{tabular}{l|c|c|c}
		\textbf{Condition} & \textbf{Amplitude~[nA]} & \textbf{$\mu$ [eV]} & \textbf{$\sigma$ [eV]} \\\hline
		2374 before & -20.02 & 150.35 & 21.21 \\
		2374 after  & -21.89 & 154.90 & 27.83 \\
		2574 before & -19.62 & 150.95 & 24.59 \\
		2574 after  & -17.28 & 149.17 & 21.05 \\
	\end{tabular}
	\label{tab:fits}
\end{table*}
\newpage
\section*{Pre-processing}

\begin{figure}[b!]
	\centering
	\includegraphics[width=0.7\textwidth]{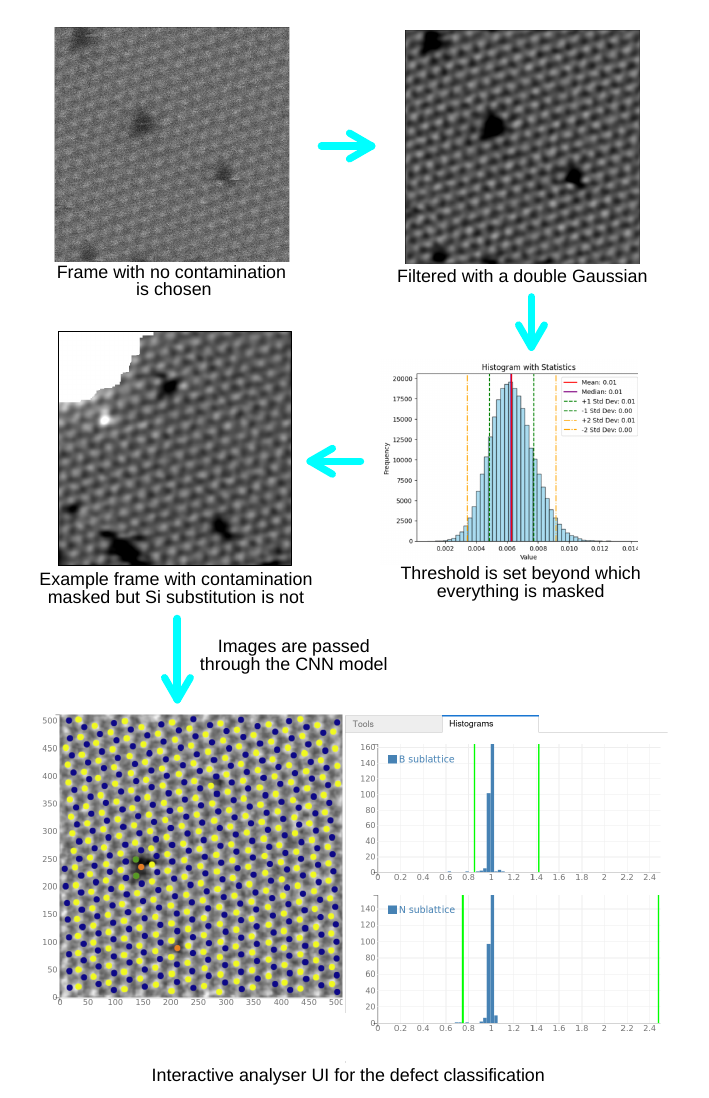}
	\caption{{\bf Illustration of the analysis workflow.}
		The images are pre-processed by filtering them with a double Gaussian.
		Contamination in the frames are then masked out by setting a threshold for what is the hBN lattice and what is contamination.
		All the frames in a stack get masked using this threshold and are then given to the convolutional neural network (CNN) for lattice identification.
		The bottom image is a screenshot of the interactive analyser user-interface where the defect classification is performed.
		The plots show histograms of the boron and nitrogen sublattices, and the vertical green bars are moved to set the thresholds.
	}
	\label{fig:pre_processing}
\end{figure}
\newpage
\section*{CNN analysis workflow}

\begin{figure}[b!]
	\centering
	\includegraphics[width=0.95\textwidth]{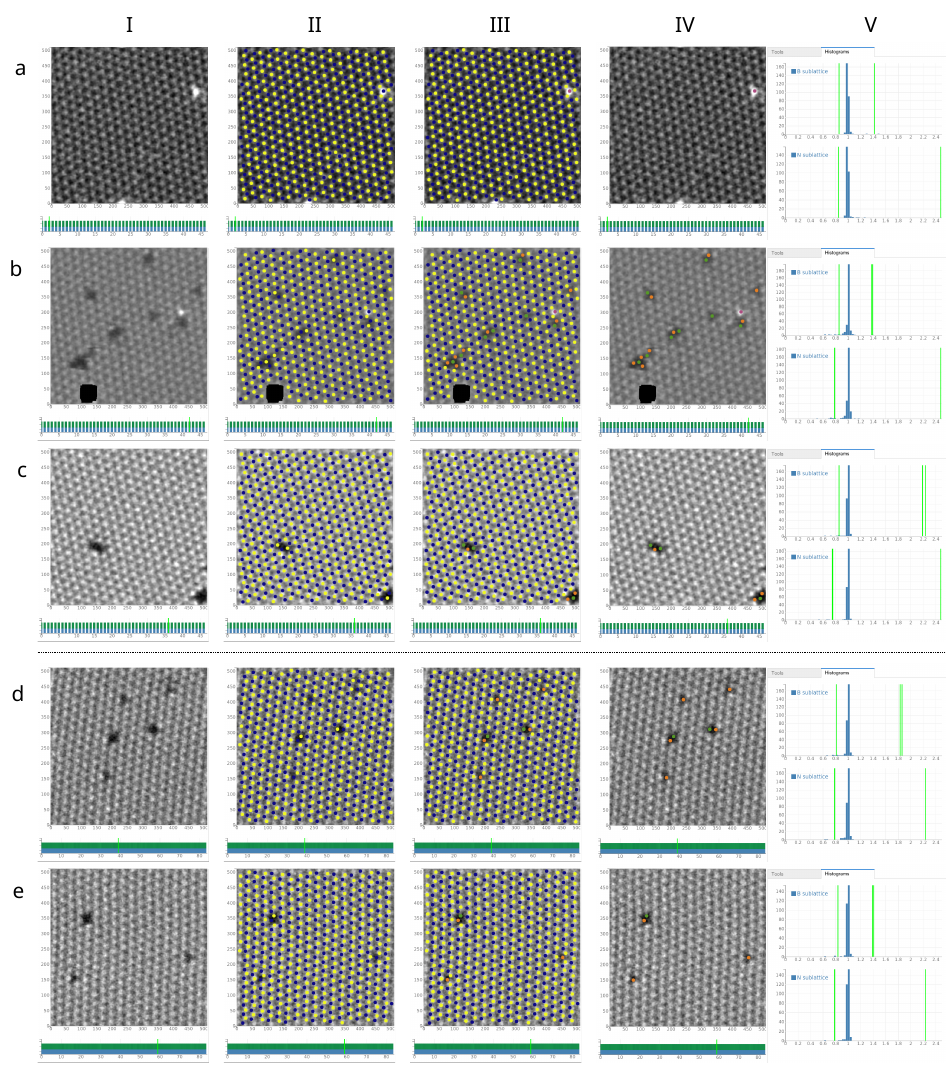}
	\caption{{\bf Identifying defects and heteroatoms by thresholding in the CNN analysis workflow.}
		In each row the same image is shown with different overlays (I-IV) and the histogram with the thresholds applied to them (V).
		Column I shows the double-Gaussian filtered image that was fed into the CNN, column II shows the CNN overlay of the identified lattice positions with blue and yellow dots marking the boron and nitrogen sublattices.
		Column III shows the defects identified via the thresholds set in column V.
		The identified defects are each shown in different colours, $\mathrm{V_B}$ being orange, $\mathrm{V_N}$ being green, and $\mathrm{Si_B}$ being purple.
	}
	\label{fig:NN_working}
\end{figure}

\clearpage
\section*{Ehrenfest simulations}

\begin{figure}[h!]	
	\centering
	\includegraphics{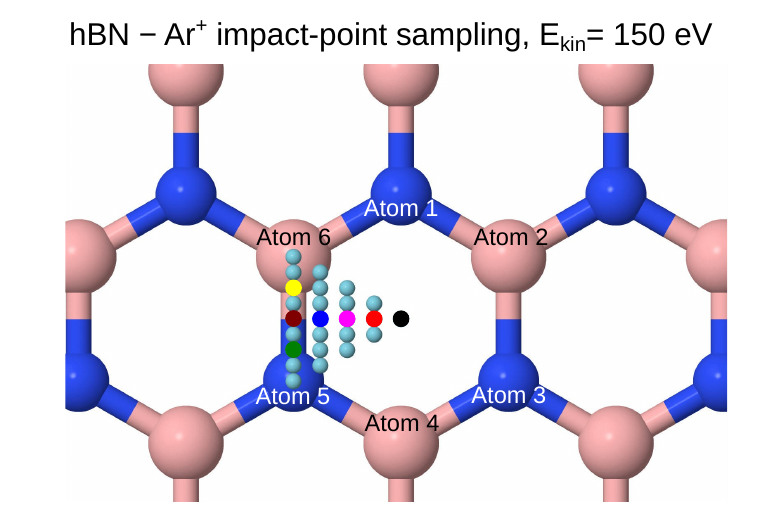}
	\caption{{\bf Ehrenfest dynamics simulation of 150~eV Ar$^+$ impinging on pristine hBN.}
		Illustration of the impact point grid (in cyan) with seven differently coloured points (black, red, pink, blue, yellow, brown and green) where Ehrenfest trajectories were calculated.
	}
	\label{fig:impact_points}
\end{figure}

\newpage
\section*{Limitations of the CNN analysis}

\begin{figure}[h!]
	\centering
	\includegraphics[width=0.85\textwidth]{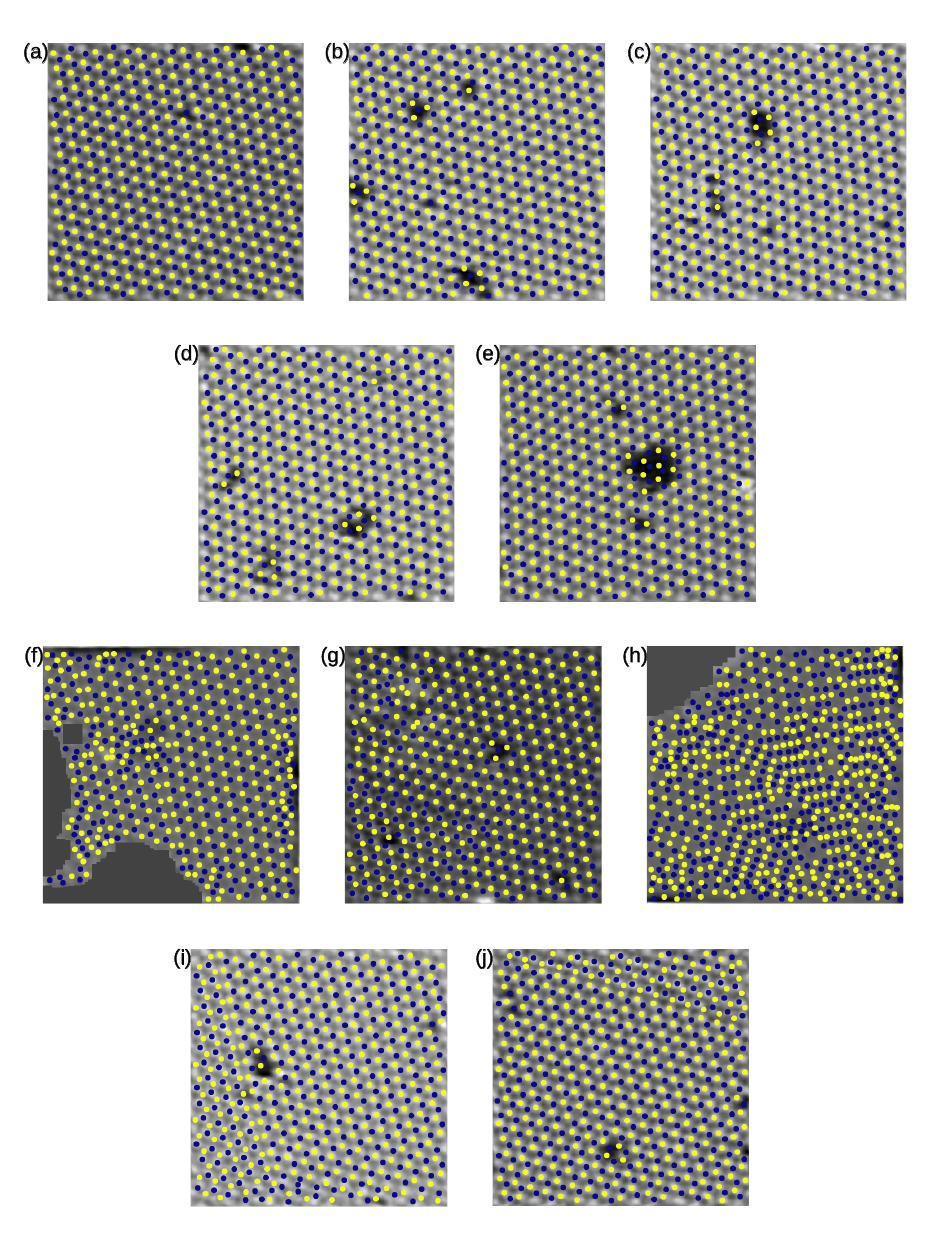}
	\caption{{\bf Example images of the lattice recognition by the CNN.}
		Panels (a) to (e) showcase the correct identification of the nitrogen and boron sublattices in yellow and blue.
		Panels (f) to (j) showcase some instances where the lattice identification fails and hence these frames had to be discarded and not analyzed.
	}
	\label{fig:lattice_recognition}
\end{figure}

\begin{figure}[h!]
	\centering
	\includegraphics{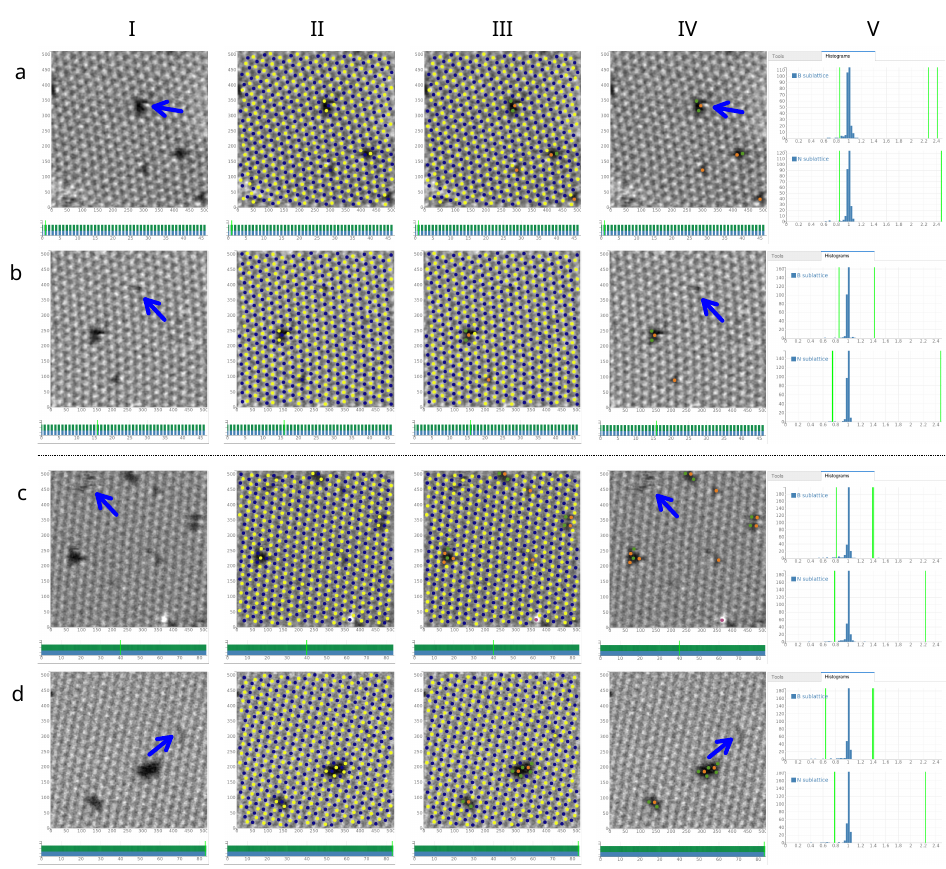}
	\caption{{\bf Limitations of identifying defects.}
		The figure follows the same logic as Fig.~\ref{fig:NN_working}.
		Blue arrows mark areas where the overlay of defects disagrees with the scattering contrast that is observed in the image due to different reasons.
		(a) A bright atom is moving during the scan and thus the average intensity at the atomic center is relatively low.
		If the threshold is lowered, however, many N positions will be misidentified as $\mathrm{Si_N}$, thus the atom is not included in the statistics.
		(b) A boron atom seems to disappear during the scan; however the threshold was not able to incorporate this as a defect without misidentifying many B atoms as $\mathrm{V_B}$.
		(c) The blurred intensity is, in all likelihood, carbon atoms moving between different vacancy positions.
		This region is not identified as a defect because including it would cause many single vacancies to be misidentified.
		(d) $\mathrm{V_B}$ is not identified due to the same thresholding issues described above.
	}
	\label{fig:NN_not_working}
\end{figure}

\clearpage

\section*{Possible oxygen substitutions}

\begin{figure}[h!]	
	\centering
	\includegraphics{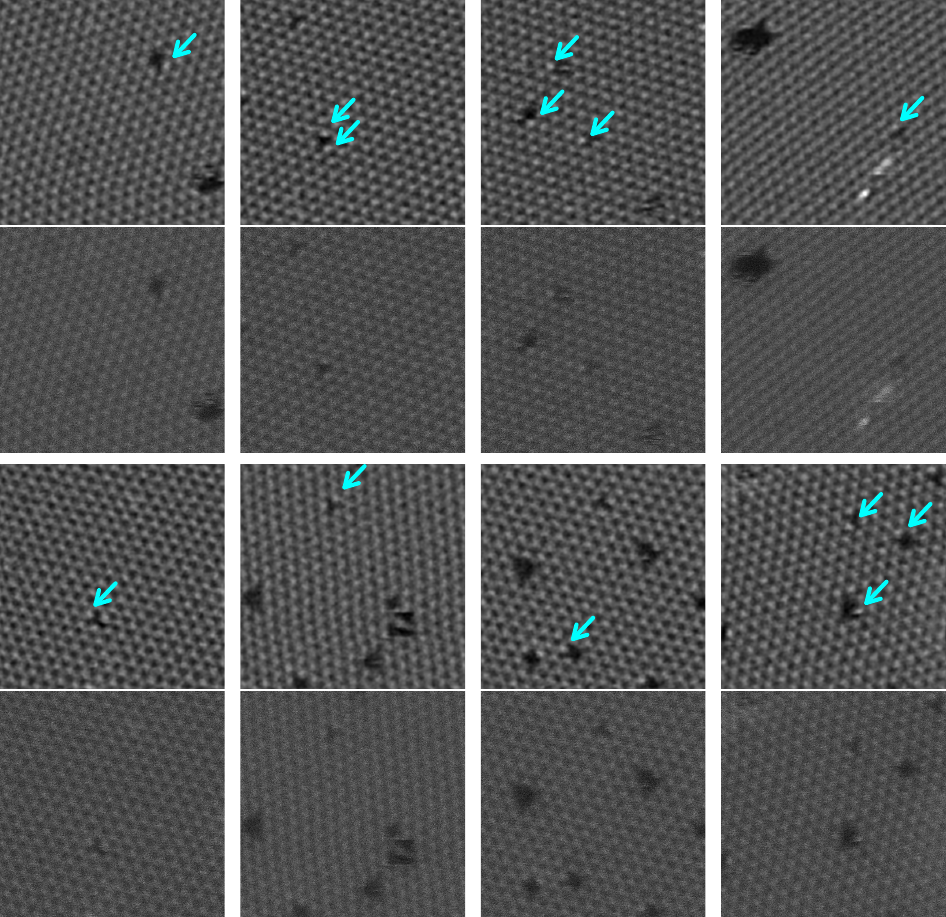}
	\caption{\textbf{Bright atoms at defect edges}.
		Double-Gaussian (top) and unfiltered (bottom) MAADF images with a FOV of $4 \times 4$~nm$^{2}$.
		Blue arrows mark atoms with higher intensities in the filtered images.
		Filter parameters: outer radius 0.12, inner radius 0.03, weight 0.3, blur 0.0.
	}
	\label{fig:exp_O_filling}
\end{figure}

\begin{figure}[h!]	
	\centering
	\includegraphics{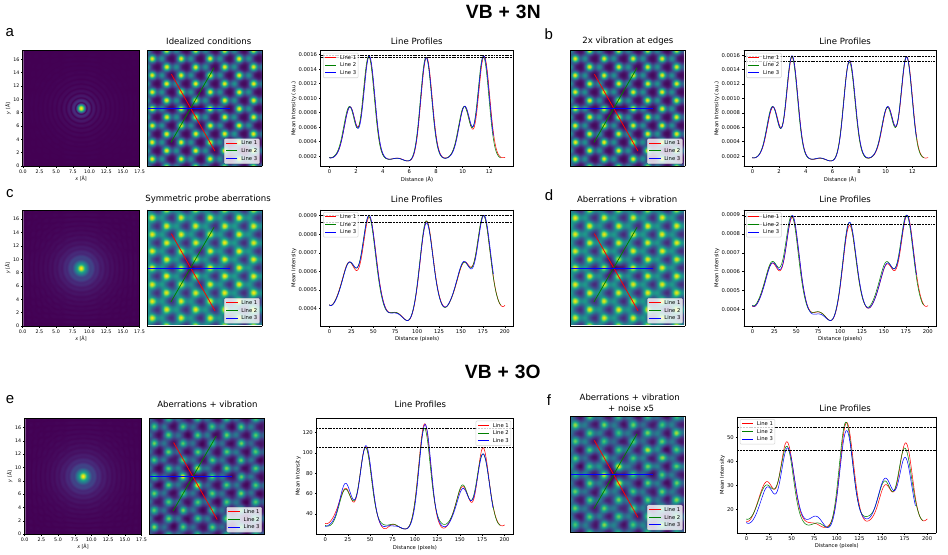}
	\caption{\textbf{Image simulations and corresponding line profiles for qualitative demonstration of intensity changes next to \boldmath$\mathrm{V_B}$}.
		(a) Idealized electron probe (in square root scale to make probe tails visible) and resulting image simulation of $\mathrm{V_B}$ surrounded by nitrogen atoms.
		(b) Equivalent simulation as in panel a but with increased standard deviation of atomic displacements for the surrounding nitrogen atoms.
		(c) Electron probe with spherical aberration  (in square root scale) and image simulation with defocus spread of the same system as in panel a.
		(d) Equivalent simulation as in panel c but with increase vibration of surrounding nitrogen atoms as in panel c.
		(e) Image simulation of oxygen atoms replacing nitrogen atoms around the $\mathrm{V_B}$ site using the same parameters and probe as in panel d.
		(f) Equivalent simulation as in panel e but with a five times increased noise level.
	}
	\label{fig:O_filling}
\end{figure}

We observed that a significant number of $\mathrm{V_B}$ defects were surrounded by atoms that appeared not only brighter than other nitrogen sites but also slightly larger in size, see Fig.~\ref{fig:exp_O_filling}.
This observation is somewhat unexpected, as atoms located next to edges or atomic defects typically exhibit lower intensities than atoms within the pristine lattice.
Such a reduction in intensity is mainly attributed to two effects:
First, the electron probe does not consist solely of a bright central maximum but also exhibits so-called probe tails, which interact with the atomic potential beyond the immediate probe position. 
These probe tails arise due to several physical effects, including finite convergence angle–related Airy disks, as well as spherical and higher-order aberrations.
In practice, this means that the ADF detector records scattering not only from the atom directly at the probe position but also from neighboring atoms.
Atoms located next to edges or vacancies have fewer neighboring atoms, leading to reduced scattering and consequently lower detected intensities.
Second, atoms at edges or vacancies are under-coordinated and therefore can move more freely in comparison to atoms within the pristine lattice, resulting in a more smeared-out appearance and lower measured intensity in the ADF-STEM images.
This is reflected in the reduced displacement threshold energy of nitrogen atoms at $\mathrm{V_B}$ edges (15.00~eV) compared to that in pristine regions (23.06~eV)~\cite{kotakoski_Electron_2010}.
In addition, kernel-based image filtering routines such as Gaussian blurring can further reduce the apparent intensity of atoms adjacent to vacancies, as these positions receive less intensity transfer from neighboring sites.

Despite these considerations, we frequently observe similar or even higher intensities next to $\mathrm{V_B}$ defects.
This suggests that the surrounding nitrogen atoms have been replaced by oxygen atoms.
Oxygen is the most plausible candidate for the observed intensities, as it is only slightly heavier than nitrogen, is present in hydrocarbon contamination~\cite{leuthner_Scanning_2019}, and is chemically more likely, and far more abundant, to substitute nitrogen than neighbouring elements such as fluorine.
In Fig.~\ref{fig:O_filling}, we illustrate this reasoning using multislice \textit{ab}TEM image simulations.
Fig.~\ref{fig:O_filling}a shows a simulated image of a $\mathrm{V_B}$ defect surrounded by nitrogen atoms under idealized conditions, where the probe is described by an Airy disk corresponding to a semi-convergence angle of 35~mrad.
The corresponding line profiles show only a negligible decrease in intensity at the nitrogen sites next to the $\mathrm{V_B}$.

In Fig.~\ref{fig:O_filling}b, we present the same simulation, but this time the smearing of intensity is modeled by doubling the standard deviation of atomic displacements for the surrounding nitrogen atoms.
This results in a significant decrease in intensity.
We emphasize that this model should be regarded as a qualitative demonstration, as the frozen phonon approach—while accurately describing thermal diffuse scattering—does not fully capture the smeared intensity.
Fig.~\ref{fig:O_filling}c illustrates the impact of spherical and chromatic aberrations as described in the Methods section.
The resulting probe exhibits substantially larger probe tails, leading to a pronounced reduction in the intensity of the surrounding nitrogen atoms.
This decrease becomes even more pronounced when both aberrations and increased vibrations are combined, as shown in Fig.~\ref{fig:O_filling}d.
We note that non-radially symmetric aberrations may have an even stronger influence on relative intensities, but due to their possible coupling with the atomic potential~\cite{lamprecht_Uncovering_2025}, their effects are not straightforward to describe in an unbiased manner.

Finally, Fig.~\ref{fig:O_filling}e demonstrates that even when both vibrational and radially symmetric aberration effects are included, oxygen substitutions around the $\mathrm{V_B}$ defect are expected to exhibit higher intensities than nitrogen sites in the pristine lattice.
Because the experimental images analyzed in this study are typically much noisier than the simulated ones, we show in Fig.~\ref{fig:O_filling}f the same simulation including oxygen substitutions, but with the Poisson-distributed noise increased by a factor of five.
Even in this case, the oxygen-substituted sites maintain at least comparable intensity to nitrogen sites, qualitatively supporting our interpretation that the bright atoms adjacent to $\mathrm{V_B}$ correspond to oxygen replacements or filling of larger defect structures.

\end{document}